\documentclass[twocolumn,aps,pra,showpacs]{revtex4-1}

\usepackage{graphicx}
\usepackage{graphics}
\usepackage{epsfig}
\usepackage{latexsym}
\usepackage{amsmath}
\usepackage{amssymb}

\def\endproof{\vrule height6pt width6pt depth0pt}

\begin{document}


\title{Proposed test of macroscopic quantum
contextuality}


\author{Ad\'{a}n Cabello}
\affiliation{Departamento de F\'{\i}sica Aplicada II,
Universidad de Sevilla, E-41012 Sevilla, Spain}


\date{\today}



\begin{abstract}
We show that, for any system with a number of levels which can
be identified with $n$ qubits, there is an inequality for the
correlations between three compatible dichotomic measurements
which must be satisfied by any noncontextual theory, but is
violated by any quantum state. Remarkably, the violation grows
exponentially with $n$, and the tolerated error per correlation
also increases with $n$, showing that state-independent quantum
contextuality is experimentally observable in complex systems.
\end{abstract}



\maketitle


\section{Introduction}


Quantum mechanics (QM) is universal, that is, applicable to any
physical system on which experiments can be made. However, a
widely held opinion is that the kinds of phenomena that makes
QM striking, like the nonexistence of definite outcomes and the
need of superpositions to describe the state of a system, are
relevant only for relatively simple ``microscopic'' systems.
The more complex the system is, the greater the effect of noise
and decoherence; thus these quantum phenomena soon become
unobservable. In this paper we describe how one of these
phenomena, quantum contextuality, is ``easier'' to observe in a
complex system than in a simple one.

Quantum contextuality, that is, the impossibility of
reproducing QM with noncontextual hidden variable (HV) models
in which observables have preassigned results which are
independent of any compatible measurements being carried out on
the same system, was independently discovered by Kochen and
Specker (KS) \cite{Specker60, KS67}, and Bell \cite{Bell66}.
Noncontextuality is motivated by the observation that, when two
compatible observables are sequentially measured any number of
times on the same system, their results do not change. Locality
is a particular form of noncontextuality supported by the
assumption that the result of a measurement does not depend on
spacelike separated events. An interesting feature of quantum
contextuality is that, while quantum nonlocality is specific
for some states of composite systems, quantum contextuality can
be observed for any state of any system with more than two
compatible observables. Indeed, state-independent quantum
contextuality (SIQC) is experimentally testable
\cite{Cabello08, BBCP09}. So far, it has only been observed in
two-qubit systems: pairs of ions with two internal levels
\cite{KZGKGCBR09}, polarization and path of single photons
\cite{ARBC09}, and two-qubit nuclear magnetic resonance systems
\cite{MRCL09}. Related quantum contextuality experiments for
specific states have been performed using also two-qubit
systems: spin and path of single neutrons \cite{BKSSCRH09}, and
polarization of two-photon systems \cite{LHGSZLG09}. A natural
question is whether SIQC can be observed in larger physical
systems. In order to answer it, in this paper we address two
problems: (a) Given an $n$-qubit system, and for three
sequential measurements, is there a better inequality to
observe SIQC, in the sense of providing a violation more
resistant against experimental imperfections? (b) How does the
robustness against imperfections scale with $n$?


\section{Resistance to imperfections}


We will focus on inequalities of the form \cite{Cabello08,
KZGKGCBR09, ARBC09, MRCL09}
\begin{equation}
 \chi :=
 \sum_{i=1}^S \langle {\cal C}_i \rangle
 - \sum_{i=S+1}^N \langle {\cal C'}_i \rangle
 \le b,
 \label{inequality}
\end{equation}
where $\langle {\cal C}_i \rangle$ and $\langle {\cal C'}_i
\rangle$ are the mean values of the product of three compatible
observables measured sequentially. A set of compatible
observables is called a context. An experimental test of
\eqref{inequality} requires testing $N$ contexts: $S$ of them
are ``positive'' contexts, defined as those in which the
product of the operators which represent the observables is the
identity $\openone$ (therefore, according to QM, the product of
the outcomes must be $1$); $N-S$ of them are ``negative''
contexts, defined as those in which the product is $-\openone$
(therefore, the product of the outcomes must be $-1$). $b$ is
the upper bound of $\chi$ in any noncontextual HV theory; it
can be obtained by examining all possible combinations of
noncontextual outcomes. Alternatively, it can be easily seen
that $b=2 s-N$, where $s$ is the maximum number of quantum
predictions which can be simultaneously satisfied by a
noncontextual HV model. For instance, in the inequality tested
in \cite{KZGKGCBR09, ARBC09, MRCL09}, there are $N=6$
predictions and a noncontextual HV model can only satisfy $s=5$
of them at most. Therefore, the bound is $b=4$. The quantum
prediction (assuming an ideal experiment without imperfections)
maximally violates inequality \eqref{inequality}, that is,
$\chi_{\rm QM} = N$. In practice, the experimental values are
in the range $\chi_{\rm expt}=5.2$--$5.5$ \cite{KZGKGCBR09,
ARBC09, MRCL09}, instead of $\chi_{\rm QM} = 6$, since the
experimental values for the correlations are $|\langle {\cal
C}_i \rangle| =1-\epsilon_i$, with $0 < \epsilon_i \ll 2$,
instead of the quantum predictions for an ideal experiment,
$|\langle {\cal C}_i \rangle| =1$; therefore, $\chi_{\rm QM}
\rightarrow \chi_{\rm expt}= \chi_{\rm QM} - \sum_{i=1}^N
\epsilon_i$. The experiments do not reach $\chi_{\rm QM}$ for
different reasons, for example, nonperfect unitary operations
and entangling gates \cite{KZGKGCBR09}, and nonperfect
alignment of the interferometric setups \cite{ARBC09}.

Experimental imperfections can also be interpreted as a failure
of the assumption of perfect compatibility under which the
bound $b$ is valid, and force us to correct this bound. This
correction takes the form $b \rightarrow b'= b + \sum_{i=1}^N
\phi_i$, where $\phi_i > 0$ can be obtained from additional
experiments \cite{GKCLKZGR10}.

Assuming that all correlations are affected by similar errors,
i.e., that $\sum_{i=1}^N \epsilon_i = N \epsilon$ and
$\sum_{i=1}^N \phi_i = N \phi$, we can define the error per
correlation as $\varepsilon= \epsilon+\phi$. A natural measure
of robustness of a quantum violation of the inequality
\eqref{inequality} against imperfections is the tolerated error
per correlation, which can be expressed as
\begin{equation}
 \varepsilon = \frac{\chi_{\rm QM}-b}{N}.
\end{equation}
If $\phi$ is negligible, then $\varepsilon$ is the maximum
difference that can be tolerated (still violating the
inequality) between the experimental value of a correlation and
the quantum value for an ideal experiment. In general, $\phi$
is not negligible, but is similar for experiments with
sequential measurements of the same length, then $\varepsilon$
is a good measure to compare the resistance to imperfections of
inequalities involving correlations between the same number of
measurements. A different argument supporting this conclusion
can be found in \cite{KGPLC10}.

The inequality tested in \cite{KZGKGCBR09, ARBC09, MRCL09}
involves correlations between sequential measurements of three
observables on two-qubit systems and tolerates an error per
correlation of $\varepsilon=1/3 \approx 0.33$. However, none of
these experiments on SIQC is free of the compatibility loophole
\cite{GKCLKZGR10}. With the imperfections in the sequential
measurements of these experiments we would need an inequality
tolerating $\varepsilon \approx 0.48$. The fact that some
experiments using specific states are free of this loophole
\cite{KZGKGCBR09} suggests that there is no fundamental reason
why we could not perform a loophole-free experiment on SIQC.
One way to face the problem is to improve the experimental
techniques. Another approach is to find inequalities with a
higher degree of robustness against imperfections. Previously
proposed inequalities are not good enough for this purpose. For
instance, the inequality based on the proof of the KS theorem
in dimension $3$ with the fewest number of contexts
\cite{Peres91} tolerates $\varepsilon= 2/17 \approx 0.12$. This
explains why it is not reasonable to expect conclusive
experimental violations of any of these inequalities and raises
the problem of whether these hypothetical more robust
inequalities do actually exist. A different approach to deal
with the loopholes created by some of the assumptions in which
inequality is based (perfect compatibility between sequential
measurements or one-to-one correspondence between actual
measurements and projective measurements) is to combine
sequential measurements on a single system with additional
measurements on a highly correlated distant ancillary system
\cite{Cabello10}.


\section{Optimal two-qubit inequality}


We will first construct a more robust inequality for two
qubits, and then prove that this inequality is the one with the
highest $\varepsilon$. For inequalities of the form
\eqref{inequality}, $\varepsilon = 2 (N-s)/N$. Therefore, to
obtain a more robust inequality, we must increase the ratio of
predictions which cannot be satisfied with a noncontextual HV
model. This can be done as follows. Let us start with the proof
of the KS theorem in Table~\ref{Table1}. It has nine
observables and $N=6$ predictions, but only $s=5$ of them can
be reproduced with a noncontextual HV model. Therefore, the
corresponding inequality is
\begin{equation}
 \sum_{i=1}^3 \langle R_i \rangle - \sum_{j=1}^3 \langle C_j \rangle
 \le 4,
\end{equation}
and tolerates $\varepsilon=\frac{1}{3} \approx 0.33$, since the
quantum prediction is $6$.


\begin{table}[htb]
\caption{\label{Table1}Proof of the KS theorem for two qubits.
$X_1 Y_2=\sigma_x^{(1)} \otimes \sigma_y^{(2)}$. Each row $R_i$
contains a positive context and each column $C_i$ a negative
one. It is impossible to assign predefined noncontextual
outcomes ($-1$ or $+1$) to the nine observables and satisfy the
six predictions of QM.}
\begin{ruledtabular}
\begin{tabular}{c|ccc|c}
 & $C_1:=$ & $C_2:=$ & $C_3:=$ & $\prod$ \\
 \hline
 $R_1:=$ & $X_1 X_2$ & $Y_1 Z_2$ & $Z_1 Y_2$ & $=\openone$ \\
 $R_2:=$ & $Y_1 Y_2$ & $Z_1 X_2$ & $X_1 Z_2$ & $=\openone$ \\
 $R_3:=$ & $Z_1 Z_2$ & $X_1 Y_2$ & $Y_1 X_2$ & $=\openone$ \\
 \hline
 $\prod$ & $=-\openone$ & $=-\openone$ & $=-\openone$ & \\
\end{tabular}
\end{ruledtabular}
\end{table}


If we preserve one row and one column of Table~\ref{Table1}, we
can construct a proof of the KS theorem by adding single-qubit
observables. For example, preserving $R_2$ and $C_1$, we obtain
the proof in Table~\ref{Table2} \cite{Peres90, Mermin90b},
which leads to the inequality tested in \cite{KZGKGCBR09,
ARBC09, MRCL09}.


\begin{table}[htb]
\caption{\label{Table2}Proof of the KS theorem tested in
\cite{KZGKGCBR09, ARBC09, MRCL09}.}
\begin{ruledtabular}
\begin{tabular}{c|ccc|c}
 & $C_1:=$ & $L_{ZX}:=$ & $L_{XZ}:=$ & $\prod$ \\
 \hline
 $L_{XX}:=$ & $X_1 X_2$ & $X_2$ & $X_1$ & $=\openone$ \\
 $R_2:=$ & $Y_1 Y_2$ & $Z_1 X_2$ & $X_1 Z_2$ & $=\openone$ \\
 $L_{ZZ}:=$ & $Z_1 Z_2$ & $Z_1$ & $Z_2$ & $=\openone$ \\
 \hline
 $\prod$ & $=-\openone$ & $=\openone$ & $=\openone$ & \\
\end{tabular}
\end{ruledtabular}
\end{table}


From Table~\ref{Table1} we can obtain eight other tables by
keeping a row and a column and adding single-qubit observables.
If we consider the resulting ten tables (Table~\ref{Table1}
plus the nine ones like Table~\ref{Table2}), we have a proof of
the KS theorem involving $15$ observables, $15$ contexts, and
ten critical proofs of the KS theorem. The point is that, in
this proof the maximum number of predictions that can be
simultaneously satisfied using a noncontextual HV model is $12$
out of $15$. Therefore, if we consider the inequality involving
all $15$ contexts,
\begin{equation}
\sum_{i=1}^3 \langle R_i \rangle + \sum_{p,q \in \{X,Y,Z\}} \langle
L_{pq} \rangle - \sum_{j=1}^3 \langle C_j \rangle \le 9,
\label{RioNegroinequality}
\end{equation}
then, $\varepsilon=0.4$, since the quantum prediction is $15$.
Indeed, by checking all possibilities, it can be seen that
\eqref{RioNegroinequality} is the inequality which tolerates
the maximum amount of errors, under the assumptions that only
sequences of three measurements are considered, that these
measurements are of the form $\sigma_i^{(1)} \otimes
\sigma_j^{(2)}$, where $i,j=0,x,y,z$ and $\sigma_0=\openone$,
and that all the correlations appear with the same weight.


\section{Inequality for $n$-qubit systems}


How does the resistance to imperfection scale with the number
of qubits when the experiments are limited to sequences of {\em
three} measurements? The natural way to scale up inequality
\eqref{RioNegroinequality} to a system of $n>2$ qubits is by
considering all $n$-qubit observables represented by $n$-fold
tensor products of the form $M_{1}\otimes \cdots \otimes
M_{n}$, where $M_{i}$ is either the $2 \times 2$ identity
matrix $I=\sigma_0$ or one of the Pauli matrices
$X=\sigma_{x}$, $Y=\sigma_{y}$, and $Z=\sigma_{z}$. For a given
$n$, there are $4^n-1$ observables. For simplicity's sake, we
will use the following notation: $IXYZ=\sigma_0 \otimes
\sigma_x \otimes \sigma_y \otimes \sigma_z$. We also need to
consider all possible contexts such that the product of three
compatible observables is $\openone$ (positive contexts) or
$-\openone$ (negative contexts). Hereafter, by ``observables''
and ``contexts'' we will mean these kinds of observables and
contexts.


{\em Lemma~1: }For a given $n$, the total number of contexts is
\begin{equation}
N(n)=\frac{1}{3}(4^n-1)(4^{n-1}-1).
\end{equation}


{\em Proof: }$N(n)$ is equal to the number of pairs of
compatible observables $P(n)$ divided by~3 (since there are
three pairs in every trio). There are $4^n-1$ observables and
each of them is compatible with $2 (4^{n-1}-1)$ observables.
This can be proven as follows: two observables are compatible
if and only if they have an even number (including zero) of
qubits in which both have different Pauli matrices. For
instance, $XXX$ is compatible with $XXI$ (since they have zero
qubits with different Pauli matrices), and with $XYZ$ (since
they have two qubits with different Pauli matrices). Given an
observable $O$, half of the $4^n$ observables (including
$\openone$) have an even number (including zero) of qubits with
different Pauli matrices. Two of them are $O$ itself and
$\openone$. Therefore, there are $\frac{4^n}{2}-2$ compatible
observables with $O$. Therefore,
$P(n)=(4^n-1)(4^{n-1}-1)$.\hfill\endproof


{\em Lemma~2: }For a given $n$, the number of negative contexts
is
\begin{eqnarray}
N(n)-S(n) & = &
 \frac{1}{6}
 \sum_{c=0}^{n-2}
 \sum_{a,b}
 {n \choose c} {n-c \choose a} {n-c-a \choose b}
 \nonumber \\
 & & \times 3^{2n-a-b-2c},
\label{negatives}
\end{eqnarray}
where $a, b \ge 0$, $a+b$ is even, $\lfloor \frac{a}{2} \rfloor
+ \lfloor \frac{b}{2} \rfloor$ (where $\lfloor x \rfloor$
denotes the greatest integer less than or equal to $x$) is odd,
and the sum extends to all $a$ and $b$ such that $a+b+c\le n$.


{\em Proof: }We can represent each negative context by a
three-row $n$-column table in which each row represents an
$n$-qubit observable, and column $i$ contains qubit $i$'s Pauli
matrices of the three observables. Then, each column belongs to
one of the following classes: It has (i) three $I$'s, (ii) one
Pauli matrix and two $I$'s, (iii) two identical Pauli matrices
and one $I$, (iv) three different Pauli matrices in ``counter
clockwise'' order (i.e., $X \rightarrow Y \rightarrow Z$), or
(v) three different Pauli matrices in ``clockwise'' order
(i.e., $X \rightarrow Z \rightarrow Y$). We will denote by $a$,
$b$, $c$, and $d$ the number of columns of the type (v), (iv),
(i), and (ii), respectively. If the three observables are
compatible, $a+b$ must be even. In order to form a negative
context, $d=0$ and $\lfloor \frac{a}{2} \rfloor + \lfloor
\frac{b}{2} \rfloor$ must be odd. Then, Eq.~\eqref{negatives}
is the result of counting all possible tables with these
restrictions. The factor $1/6$ is due to the fact that six
different tables represent the same context. Factors $3^a$ and
$3^b$ are due to the fact that there are three columns of the
types (v) and (iv), and factor $3^{2(n-a-b-c)}$ is due to the
fact that there are nine columns of the type
(iii).\hfill\endproof


{\em Lemma~3: }For a given $n$, any noncontextual HV theory
must satisfy
\begin{equation}
 \Xi (n):=
 \sum_{i=1}^{S(n)} \langle {\cal C}_i \rangle
 - \sum_{i=S(n)+1}^{N(n)} \langle {\cal C'}_i \rangle
 \le 2 S(n)-N(n).
 \label{inequalityn}
\end{equation}


{\em Proof: }The bound is two times the maximum number of
predictions that a (noncontextual HV) model can reproduce minus
the total number of predictions. The quantum predictions
corresponding to the $S(n)$ positive contexts can be reproduced
with a model in which all the outcomes are $+1$. However, this
model fails to reproduce the predictions for the $N(n)-S(n)$
negative contexts. Indeed, $S(n)$ is the maximum number of
predictions that a model can reproduce: For $n=2$, the
$N(2)=15$ contexts can be grouped in ten tables like
Tables~\ref{Table1} and \ref{Table2}. Each context appears in
four of them, and none of the tables admit a model. In any
model which reproduces $S(2)=12$ predictions, any prediction
which is not reproduced by the model is in one of these tables
together with five predictions which are reproduced by the
model. Therefore, no model can reproduce more predictions. For
$n>2$, the $N(n)$ predictions can also be grouped in tables
like Tables~\ref{Table1} and \ref{Table2}, such that none of
these tables admit a model. Any positive (negative) context and
any table in $n>2$ can be univocally obtained from one of the
$12$ positive (three negative) contexts and ten tables in
$n=2$. Similarly, any model in $n>2$ which satisfies $S(n)$
predictions can be obtained from a model in $n=2$ which
satisfies $S(2)=12$ predictions. In any of these models for
$n>2$, any prediction which is not reproduced by the model
belongs to a table with five predictions which are reproduced
by the model. Therefore, no model can reproduce more
predictions.\hfill\endproof

\begin{figure}
\centerline{\includegraphics[width=7.5cm]{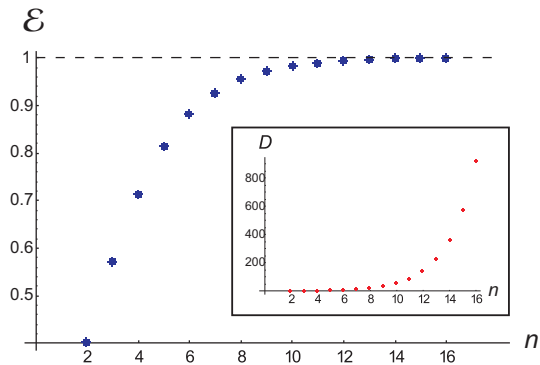}}
\caption{\label{varepsilon} Tolerated error per correlation
(still violating the inequality),
$\varepsilon$, and degree of violation, $D$, of the inequality
\eqref{inequalityn}, as a function of the number of qubits, $n$.}
\end{figure}

According to QM, any state of $n$ qubits violates inequality
\eqref{inequalityn} by the same amount, $\Xi_{\rm QM}
(n)=N(n)$. The remarkable feature is that the degree of
violation defined as the ratio between the quantum prediction
and the bound for noncontextual HV theories,
$D(n)=N(n)/[2S(n)-N(n)]$, grows {\em exponentially} with the
number of qubits $n$ (see Fig.~\ref{varepsilon}). Similarly,
the tolerated error per correlation goes as
\begin{equation}
 \varepsilon (n) = \frac{2\left[N(n)- S(n)\right]}{N(n)},
\end{equation}
which is $\frac{2}{5}=0.4$ for $n=2$, $\frac{4}{7}\approx0.57$
for $n=3$, $\frac{424}{595} \approx 0.71$ for $n=4$,
$\frac{4720}{5797} \approx 0.81$ for $n=5$, and approaches $1$
as $n$ approaches infinity (see Fig.~\ref{varepsilon}). This
happens because the ratio between the maximum number of quantum
predictions which can be satisfied by a noncontextual HV model
and the number of quantum predictions approaches $\frac{1}{2}$
as $n$ approaches infinity. Indeed, if we eliminate any context
in \eqref{inequalityn}, then the resulting inequality presents
a lower resistance to noise.


\section{Experimental implications}


A better choice for a loophole-free experiment of SIQC on a
two-qubit system than the inequality tested in previous
experiments \cite{KZGKGCBR09, ARBC09, MRCL09} is inequality
\eqref{RioNegroinequality}. This inequality requires to test
$15$ correlations, something which is feasible with the same
techniques used in previous experiments.

For a given $n$, any $n$-qubit quantum state maximally violates
\eqref{inequalityn}, and the most interesting prediction is
that the violation grows exponentially with $n$, and the
tolerated error per correlation also grows with $n$. An
experiment of inequality \eqref{inequalityn} for $n=3$,
requires to test $315$ correlations, something which is within
the range of what is experimentally feasible. Moreover, if we
choose a system in which all three qubits are equivalent in the
sense that the error in a measurement involving only qubits 1
and 2 is similar to the error in a similar measurement on any
other pair of qubits, and in which any experiment $ABC$ has a
similar error, regardless of whether $A$, $B$, and $C$ are $X$,
$Y$, or $Z$, then the experiment can be simplified invoking
this symmetry, since the $315$ contexts can be divided in seven
classes: (I) 27 like $\{XII, IXI, XXI\}$, (II) 81 like $\{XII,
IXX, XXX\}$, (III) nine like $\{XXI, YYI, ZZI\}$, (IV) 27 like
$\{XXI, XIX, IXX\}$, (V) 81 like $\{XXI, YZX, ZYX\}$, (VI) nine
negative like $\{XXI, YYI, ZZI\}$, and (VII) 81 negative like
$\{XXI, YYX, ZZX\}$, and a good estimation of the experimental
violation of the inequality can be obtained by testing a few
contexts of each class. A similar approach can be followed for
$n=4$. Actually, the most interesting experiment would be to
test the exponentially-growing-with-size violation by testing
the inequality $\eqref{inequality}$ considering an increasing
number of qubits on the same physical system. Such a test seems
feasible with current technology.


\section{Conclusions}


SIQC is universal in the sense that there is always an
inequality valid for noncontextual HV models which is violated
for any state of the system \cite{BBCP09}. However, so far, for
all previously known inequalities, the violation became smaller
as the system became more complex, making the violation
unobservable even for relatively simple systems, and suggesting
that SIQC would be observable only on very simple systems, and
that a loophole-free experiment on SIQC would not be feasible
with current technology. We have shown that, on the contrary,
the violation can increase along with the size of the system,
making it observable for larger systems.

In contrast with Bell experiments, where an exponential
violation requires correlations between $n$ separated
measurements on $n$ distant qubits \cite{Mermin90}, and the
violation decreases if the state suffers decoherence
\cite{CRV08}, the exponential violation of inequality
\eqref{inequalityn} requires only correlations between three
measurements, and is robust against decoherence.


\section*{Acknowledgments}


A.~Zeilinger was the first one who asked for a more robust
state-independent violation of noncontextual inequalities, and
O.~G{\"u}hne the first one who asked about the scaling
properties of the effect. I would also like to thank
E.~Amselem, P.~Badzi{\c a}g, I.~Bengtsson, G.~Bj\"{o}rk,
M.~Bourennane, J.-{\AA}.~Larsson, and M.~R{\aa}dmark, for
enlightening discussions, and D.~Rodr\'{i}guez for checking
Eq.~\eqref{negatives}. The author acknowledges support from the
MCI Project No.~FIS2008-05596, and the Junta de Andaluc\'{\i}a
Excellence Project No.~P06-FQM-02243.



\end{document}